\documentclass[reprint,aip,jcp]{revtex4-1}

\usepackage{hyperref}
\usepackage{amsmath}
\usepackage{amsfonts}
\usepackage{amssymb}
\usepackage{xspace}
\usepackage{graphicx}
\usepackage{multirow}
\usepackage{mathtools}
\usepackage{url}
\usepackage{setspace}
\usepackage{xparse}
\usepackage{algorithmicx}
\usepackage{algpseudocode}
\usepackage{subcaption}

\usepackage[size=small]{caption}
\usepackage{etoolbox}
\AtEndEnvironment{algorithm}{\noindent\hrulefill\par\nobreak\vskip-5pt}
\usepackage{newfloat}

\DeclareFloatingEnvironment[
    fileext=loa,
    listname=List of Algorithms,
    name=Algorithm,
    placement=tbhp,
]{algorithm}
\DeclareCaptionFormat{algorithms}{\vskip-15pt\hrulefill\par#1#2#3\vskip-6pt\hrulefill}
\usepackage{tikz}

\makeatletter
\def\dbar{{\mkern4mu\mathchar'26\mkern-13mud}}
\def\bbar{{\mathchar'26\mkern-9mub}}
\makeatother

\usetikzlibrary{positioning}
\usetikzlibrary{calc}
\usetikzlibrary{decorations.pathmorphing}

\DeclareMathOperator*{\FrobNorm}{Frob}
\DeclareMathAlphabet{\mathpzc}{OT1}{pzc}{m}{it}

\DeclareMathOperator{\ext}{ext}

\newcommand{\bigOh}{\ensuremath{O}}
\newcommand{\bigOhNone}{\ensuremath{\bigOh(N)}\xspace}
\newcommand{\bigOhN}[1]{\ensuremath{\bigOh(N^{#1})}\xspace}

\newcommand{\gbk}[2]{\ensuremath{\left( #1 \middle| #2 \right)}}

\newcommand{\si}{\ensuremath{\sigma}\xspace}
\newcommand{\la}{\ensuremath{\lambda}\xspace}
\newcommand{\erig}{\ensuremath{\mathbf{g}}\xspace}

\newcommand{\exK}{\ensuremath{\mathbf K}\xspace}
\newcommand{\Dmat}{\ensuremath{\mathbf D}\xspace}
\newcommand{\LinK}{{\tt LinK}\xspace}
\newcommand{\Lst}[2]{\ensuremath{L_{#2}^{(#1)}}\xspace}
\newcommand{\Lstset}[1]{\ensuremath{\mathbb{L}^{(#1)}}\xspace}
\newcommand{\frob}[2]{\ensuremath{\FrobNorm_{#1}\left\{#2\right\}}\xspace}
\newcommand{\peq}{\ensuremath{\,+\!=\,}}
\newcommand{\meq}{\ensuremath{\,-\!=\,}}
\newcommand{\exB}{\ensuremath{\mathbf{B}}\xspace}
\newcommand{\sbk}[1]{\ensuremath{\gbk{#1}{#1}^{1/2}}\xspace}

\newcommand{\thws}{\ensuremath{\vartheta_{\mathrm{ws}}}\xspace}
\usepackage{todonotes}

\newcommand{\thsq}{\ensuremath{\vartheta_{\mathrm{SQ}}}\xspace}
\newcommand{\sqvl}{SQV\ensuremath{\ell}\xspace}
\newcommand{\figref}[1]{Figure~\ref{fig:#1}\xspace}

\newcommand{\secref}[1]{Section~\ref{sec:#1}\xspace}
\newcommand{\Algref}[1]{Algorithm~\ref{alg:#1}\xspace}

\DeclareDocumentEnvironment{diagram}{D[]{node distance=1cm} m D[]{}}{%
\begin{equation}\begin{tikzpicture}[node distance=3mm,baseline=(current bounding box.center)]%
  \node (pic) \bgroup%
    \begin{tikzpicture}[#1]}%
{%
  \end{tikzpicture}%
  \egroup;%
  \node (k) [left=of pic,#3] {#2};%
\end{tikzpicture}\end{equation}\noindent}%

\NewDocumentCommand{\myState}{ u{\;} }{%
    \State #1%
}
\algblockdefx[ForInLoop]{ForIn}{EndForIn}%
  [2][Unknown]{\textbf{foreach} {#1} \textbf{in} {#2} \textbf{do}}%
  [1][]{\textbf{end} loop over #1}
\algblockdefx[ForInLoop]{ForWith}{}%
  [2][Unknown]{\textbf{foreach} {#1} \textbf{with} {#2} \textbf{do}}%

\begin{document}

\title{Fast construction of the exchange operator in an atom-centered basis with concentric atomic density fitting}
\author{David~S.~Hollman}
\affiliation{Center for Computational Quantum Chemistry, University of Georgia, 1004 Cedar St., Athens, Georgia, 30602, USA}
\affiliation{Department of Chemistry, Virginia Tech, Blacksburg, Virginia 24061, USA}
\author{Henry~F.~Schaefer}
\affiliation{Center for Computational Quantum Chemistry, University of Georgia, 1004 Cedar St., Athens, Georgia, 30602, USA}
\author{Edward~F.~Valeev}
\affiliation{Department of Chemistry, Virginia Tech, Blacksburg, Virginia 24061, USA}

\begin{abstract}

A linear-scaling algorithm is presented for computing the Hartree--Fock (HF) exchange matrix using concentric atomic density fitting.  The algorithm utilizes the stronger distance dependence of the three-center electron repulsion integrals along with the rapid decay of the density matrix to accelerate the construction of the exchange matrix.  The new algorithm is tested with computations on systems with up to 1536 atoms and 15585 basis functions, the latter of which represents, to our knowledge, the largest quadruple-zeta HF computation ever performed.  Our method handles screening of high angular momentum contributions in a particularly efficient manner, allowing the use of larger basis sets for large molecules without a prohibitive increase in cost.

\end{abstract}

\maketitle

\section{Introduction}

%Self consistent field (SCF) methods are ubiquitous in {\it ab initio} quantum chemistry, either as a means of determining energies and properties of chemical systems or as a starting point for correlated computations.  Recent years have seen a significant increase the use of in SCF methods that include at least some exact exchange---namely, Hartree--Fock (HF)\cite{Hartree:1947ia} and hybrid Density Functional Theory (DFT) methods\cite{Clementi:1990kd} (including the popular B3LYP\cite{Becke:1993p1372}).
%In most cases, the rate-limiting step in these methods is the computation of the exchange matrix \exK, given by
Computation of the so-called exchange operator is often the most expensive step in electronic structure models
applicable to large systems, such as the hybrid\cite{Becke:1993p1372} Density Functional Theory (DFT) and
Hartree-Fock (HF),\cite{Hartree:1947ia} the latter of which serves as the starting point for electron correlation treatment with modern
reduced-scaling many-body methods.\cite{Riplinger:2013p034106} The exchange matrix, \exK, is given by
\begin{align}
  K_{\mu\nu} = \sum_{\la\si}D^{\la\si}\gbk{\mu\la}{\nu\si},
  \label{eq:K4c}
\end{align}
where $\mu$, $\nu$, $\la$, $\si$, \dots\  denote basis functions, \Dmat is the density matrix, and 
\begin{align}
  \gbk{\mu\la}{\nu\si} = \int\chi_{\mu}(\vec r_1)\chi_{\la}(\vec r_1)\frac1{|\vec r_1 - \vec r_2|}\chi_{\nu}(\vec r_2)\chi_{\si}(\vec r_2) d\tau_1 d\tau_2
\end{align}
are the electron repulsion integrals (ERIs) in the Mulliken ``bra-ket'' notation. Here we consider
atom-centered basis functions $\chi(\vec r)$, typically represented by (contracted) Gaussian-type orbitals.
Although $\gbk{\mu\la}{\nu\si}$ decays with the bra-ket distance $R$ only as $R^{-1}$ (i.e., slowly),
the decay of the density matrix $D^{\la\si}$ in a finite system (and, generally, any system with a nonzero band gap) is exponential with the distance between $\la$ and $\si$.\cite{Goedecker:1998if}
Therefore, as first noted by Alml\"of,\cite{AlmlofYarkony} the number of nontrivial elements of the exchange matrix \exK grows linearly with system size---that is, \exK has \bigOhNone significant elements ($N \propto$ system size). A number of linear-scaling methods for the computation of \exK have been investigated over the years, with pioneering work done in chemistry in the mid-1990s by Schwegler, Challacombe, and Head-Gordon\cite{Schwegler:1996hh,Schwegler:1997p9708} and Burant, Scuseria, and Frisch.\cite{Burant:1996cf} Several related algorithms have been developed, e.g. ONX by Challacombe et al and \LinK by Ochsenfeld, White, and Head-Gordon\cite{Ochsenfeld:1998p1663}.

The aforementioned \bigOhNone algorithms for exchange take advantage of the element-wise sparsity of the ERI and density matrix tensors in Eq.~\eqref{eq:K4c}. Some effort has gone into utilization of rank sparsity of the ERI tensor, as revealed by the multipole expansion
of the Coulomb operator,\cite{Schwegler:1999bz} by the pseudospectral and related factorizations of ERI,\cite{Friesner:1985bo,Neese:2009ca,Hohenstein:2012p044103,Parrish:2012br,Hohenstein:2012hs} or, as done here, by the density fitting approximation of ERIs.\cite{Sodt:2008p104106}
The density fitting approximation\cite{Lowdin:1953bm,Whitten:1973ju,Baerends:1973p41,Jafri:1974fb,Feyereisen:1993p359,Vahtras:1993db,Eichkorn:1995fq,Eichkorn:1997bi,Bauernschmitt:1997ef,Weigend:1998p143} (DF, also called resolution of the identity), expresses the two-center products in the bra and the ket of the ERI tensor (often called ``densities,'' not to be confused with the density matrix~\Dmat) as linear combinations of one-center functions from an auxiliary basis:
\begin{align}
  \left|\mu\nu\right) = \sum_{X} C_{\mu\nu}^X \left|X\right),
  \label{eq:dfexp}
\end{align}
where $X$, $Y$, \dots\  are indices in an auxiliary basis set.  To obtain the coefficients $C_{\mu\nu}^X$, one must left-project with the auxiliary basis to obtain a system of linear equations:
\begin{align}
  \gbk{Y}{\mu\nu} = \sum_{X} C_{\mu\nu}^X \gbk YX.
\end{align}
The solution of these equations in their full form requires \bigOhN 3 effort and is thus untenable for large molecular systems.  Moreover, substitution of the form in Eq.~\eqref{eq:dfexp} into the exchange matrix expression in Eq.~\eqref{eq:K4c} does not reduce the formal scaling of the exchange matrix build.  Consequently, significant work has been done on methods of reducing the scaling of Eq.~\eqref{eq:dfexp}, principally by localizing the fitting basis.\cite{Baerends:1973p41,Gallant:1996cf,Sodt:2006p194109,Reine:2008p104101,Sodt:2008p104106,Guerra:1998bj,Watson:2003eo,Krykunov:2009ha,Hollman:2014gc,Merlot:2013p1486}  Significantly less research has been done on the use of these localized decompositions for the exchange matrix build, Eq.~\eqref{eq:K4c}, particularly in the context of the linear scaling exchange work of the late 1990s.  Indeed, only Sodt and Head-Gordon\cite{Sodt:2008p104106} have investigated a linear scaling exchange construction with local DF.  Also, Neese and coworkers\cite{Neese:2009ca} presented a linear scaling algorithm for HF exchange in the context of their ``chain-of-spheres'' exchange (COSX) method, which borrows elements from both DF and the pseudo-spectral method of Friesner, et al.\cite{Friesner:1985bo} Herein, we present a linear scaling HF exchange algorithm and implementation using the Concentric Atomic Density Fitting (CADF) approach, based on a concept that has been around for some time\cite{Baerends:1973p41,Guerra:1998bj} and was recently revived by Merlot, et al.\cite{Merlot:2013p1486} and the present authors.\cite{Hollman:2014gc}

In \secref{theory}, we briefly review the theory behind CADF and develop a diagrammatic scheme for further discussion of linear scaling methods.  \secref{link} presents a progression of algorithms that lead to the final CADF-LinK method, and \secref{impl} discusses implementation details.  \secref{results} gives the results of some benchmark computations, and \secref{conc} contains some concluding remarks.

\section{Theoretical Background}
\label{sec:theory}

\subsection{Concentric Atomic Density Fitting}

The idea behind what we call CADF has been around since the early days of DF:\cite{Baerends:1973p41} expand a product density $\left|\mu\nu\right)$ using only auxiliary functions centered on the same atom as the composite functions $\mu$ and $\nu$.  The fitting equations can thus be written
\begin{align}
  \gbk{Y_{(ab)}}{\mu_a \nu_b} = \sum_{X\in(ab)}C_{\mu_a\nu_b}^{X_{(ab)}} \gbk{X_{(ab)}}{Y_{(ab)}},
  \label{eq:cadfcoefs}
\end{align}
where $a$, $b$, \dots\ are indices of atom centers, $\mu_a$ indicates that the function $\mu$ is centered on atom $a$, and $Y_{(ab)}$ indicates that the function $Y$ is centered on either atom $a$ or atom $b$.  In this context, the use of the simple expansion of the integral tensor
\begin{align}
  \begin{split}
  g_{\mu\nu,\la\si}&\equiv\gbk{\mu_a\nu_b}{\la_c\si_d} \\
  &\approx \sum_{X\in(ab)}\sum_{Y\in(cd)} C_{\mu_a\nu_b}^{X_{(ab)}}\gbk{X_{(ab)}}{Y_{(cd)}}C_{\la_c\si_d}^{Y_{(cd)}}
\end{split}
\end{align}
yields errors in \erig that are too large to be useful for HF theory computations.  However, in 2000 Dunlap\cite{Dunlap:2000p529} noted that this expansion is linear in the density error (unless the full, non-local Coulomb metric is used), and a more robust expansion correct to second order in the density error can be constructed. In the context of CADF, this robust expansion can be written
\begin{multline}
  \gbk{\mu_a\nu_b}{\la_c\si_d}\approx \sum_{X\in(ab)} C_{\mu_a\nu_b}^{X_{(ab)}} \gbk{X_{(ab)}}{\la_c\si_d} 
  \\
  + \sum_{Y\in(cd)}\gbk{\mu_a\nu_b}{Y_{(cd)}}C_{\la_c\si_d}^{Y_{(cd)}} 
  \\
  - \sum_{X\in(ab)}\sum_{Y\in(cd)} C_{\mu_a\nu_b}^{X_{(ab)}}\gbk{X_{(ab)}}{Y_{(cd)}}C_{\la_c\si_d}^{Y_{(cd)}}
  \label{eq:robustg}
\end{multline}
The use of this robust expansion yields reasonable errors for HF theory---on the order of 2-5 times the errors arising from conventional density fitting, in most cases.\cite{Merlot:2013p1486,Hollman:2014gc}
However, Eq.~\eqref{eq:robustg} breaks the positive semidefiniteness of the \erig tensor,\cite{Merlot:2013p1486} which can cause convergence issues in rare cases.  Merlot, et al.~and our group both suggested different approaches\cite{Merlot:2013p1486,Hollman:2014gc} to correcting this issue.
While we consider the convergence issues with CADF to be eminently solvable, to throughly test the CADF approximation
we need to be able to apply CADF-based Hartree-Fock and Density Functional Theory methods to large molecules. Thus, in this paper the 
focus is on developing a practical \bigOhNone CADF-based exchange matrix algorithm, and hence the issues of positivity are set aside herein.

\subsection{Screening of Insignificant Contributions}
Due to the rapid decay of the overlap between the Gaussian basis function functions $\mu$ and $\nu$, only \bigOhNone products
$\left|\mu\nu\right)$ result in significant ERIs.  Thus, the four-center ERI tensor \erig contains \bigOhN2 significant entries.
H{\"a}ser and Ahlrichs\cite{Haser:1989p104} noted that the Cauchy--Schwarz inequality holds for the ERI tensor, and thus this exponential decay could be exploited for prescreening using the inequality
\begin{align}
  |\gbk{\mu\nu}{\la\si}| \leq |\gbk{\mu\nu}{\mu\nu}|^{1/2}\, |\gbk{\la\si}{\la\si}|^{1/2}.
  \label{eq:schwarz}
\end{align}
Using the nomenclature of Neese and coworkers,\cite{Neese:2009ca} we say that $\mu$ forms an ``S-junction'' with $\nu$ if and only if $|\gbk{\mu\nu}{\mu\nu}|^{1/2}$ is greater than some pairing threshold $\epsilon_{S}$.

\subsubsection{Diagrammatic Notation}

To illustrate the screening of significant contributions to the exchange matrix, we introduce a simple graphical method for
showing contributing factors to the sparsity of a given tensor or tensor contraction expression.  Tensor indices are shown as vertices of a diagram. Each of these indices has a range of \bigOhNone , hence a diagram with $k$ vertices denotes to \bigOhN{k} total contributions to
a tensor or a tensor contraction. An edge connecting two vertices denotes that only \bigOhNone ~{\em pairs} of these indices are significant for any finite precision in the limit of infinite system size. In other words, for each first index value there is a $\bigOh(1)$ list of significant second index values.
For example, the ERI tensor screened using Eq.~\eqref{eq:schwarz} can be represented as
\begin{diagram}{$\gbk{\mu\la}{\nu\si}\,\,:$}
    \node (mu)               { $\mu$ };
    \node (la) [right=of mu] { $\la$ };
    \node (nu) [right=of la] { $\nu$ };
    \node (si) [right=of nu] { $\si$ };
    \draw (mu) -- node[above,font=\small] {S} (la);
    \draw (nu) -- node[above,font=\small] {S} (si);
  \label{d:g}
\end{diagram}
\hspace{-0.8em} From this diagram, it is immediately apparent that the screening in Eq.~\eqref{eq:schwarz} requires \bigOhN2 integrals to be computed for \erig, since there are a constant number of $\la$ for each $\mu$ and a constant number of $\si$ for each $\nu$, but no connection between $\mu$ and $\nu$. Therefore, the diagram must be fully connected for a tensor or contraction to have \bigOhNone ~significant contributions.

\subsubsection{Linear Scaling Exchange}

\bigOhNone\ exchange construction algorithms take advantage of the aforementioned rapid decay of the density matrix.\cite{Schwegler:1996hh,Schwegler:1997p9708,Burant:1996cf,Ochsenfeld:1998p1663} Diagrammatically, the conventional exchange matrix expression
\begin{align}
  K_{\mu\nu} = \sum_{\la\si}D^{\la\si}\gbk{\mu\la}{\nu\si}
  \label{eq:K4c2}
\end{align}
can be represented as
\begin{diagram}{$K_{\mu\nu}\,\,:$}
  \node (mu)               { $\mu$ };
  \node (la) [right=of mu] { $\la$ };
  \node (si) [right=of la] { $\si$ };
  \node (nu) [right=of si] { $\nu$ };
  \draw (mu) -- node[above,font=\small] {S} (la);
  \draw (nu) -- node[above,font=\small] {S} (si);
  \draw (la) -- node[above,font=\small] {P} (si);
  \label{eq:Kdiag}
\end{diagram}
\hspace{-0.8em} where the connection between $\la$ and $\si$ is called a ``P-junction'' in the nomenclature of Neese, et al.\cite{Neese:2009ca}
Diagram \eqref{eq:Kdiag} is fully connected, which suggests the existence of a linear scaling algorithm.
Note also that the path length between two indices in the result suggests (in a qualitative sense) some aspects of the performance of the associated linear scaling algorithm.  Since there are a bounded number of $\la$ for a given $\mu$---we will use the notation $C_{S}(\mu)$---as well as a bounded number of $\si$ given $\la$ and a bounded number of $\nu$ given $\si$, the prefactor of the linear scaling algorithm will be bounded from above by the product $C_{S}(\mu)C_{P}(\la)C_{S}(\si)$, because one could, at worst, use the direct product of these index sets to construct the tensor.  In practice, the scaling will be much better than this (since, e.g., the product of medium-sized contributions to each connection may still be quite small), though there may not necessarily be a reasonable algorithm to access this scaling.  Nevertheless, the diagrammatic approach presents a concise picture of the nature of the factors that necessarily contribute to the performance of any linear scaling algorithm for the construction of \exK.   It is clear from the form of the Gaussian product rule (GPR) that $C_{S}(\mu)$ and $C_{S}(\si)$ will depend on the number of basis functions per atom and the diffuseness of those functions.  From physical reasoning, we note that $C_{P}(\la)$ will further depend on the band gap of the system and the degree of delocalization of electrons in the system.  Given these properties of S and P junctions, we conclude that a linear scaling algorithm associated with this diagram would probably perform poorly for large basis sets or systems with small band gaps.  While this is not a particularly profound insight (both of these are well-known issues with virtually all linear scaling methods), it suggests by contrast what a method that improves on these limitations might look like: such a method would have to provide an alternative path between $\mu$ and $\nu$ that does not rely on the square of the S junction constant (for large basis sets) or the P-junction constant (for smaller band-gap systems). Our task herein is to identify useful additional pathways between $\mu$ and $\nu$.
%\dshsays{I revised this next part a little per your point.  If you still don't like it, we can take it out}

\subsubsection{CADF with Linear Scaling Exchange}

CADF introduces another kind of junction between indices, that of concentricity.  Using the diagrammatic approach, we can write the coefficient tensor as a union of two diagrams:
\begin{widetext}
\begin{diagram}[node distance=4mm]{$C_{\mu_a\la_b}^{X_{(ab)}}\,\,:$}[font=\large]
  \node (lpic) {
    \begin{tikzpicture}[node distance=1cm]
      \node (mu)  { $\mu_a$ };
      \node (la) [right=of mu] { $\la_b$ };
      \node (X) [below=of mu] { $X_a$ };
      \draw (mu) -- node[above,font=\small] (S) {S} (la);
      \draw (X) -- node[left,font=\small] {CA} (mu);
    \end{tikzpicture} 
  };
  \node (un) [font=\Large, right=of lpic] {$\cup$};
  \node (rpic) [right=of un] {
    \begin{tikzpicture}[node distance=1cm]
      \node (mu)  { $\mu_a$ };
      \node (la) [right=of mu] { $\la_b$ };
      \node (X) [below=of la] { $X_b$ };
      \draw (mu) -- node[above,font=\small] (S) {S} (la);
      \draw (X) -- node[left,font=\small] {CA} (la);
    \end{tikzpicture} 
  };
  \node (eq) [font=\Large, right=of rpic] {$=$};
  \node (epic) [right=of eq] {
    \begin{tikzpicture}[node distance=1cm]
      \node (mu)  { $\mu_a$ };
      \node (la) [right=of mu] { $\la_b$ };
      \node (mid) at ($ (mu) !0.5! (la) $) {};
      \node (X) [below=of mid.south west] { $X_{(ab)}$ };
      \draw (mu) -- node[above,font=\small] (S) {S} (la);
      \draw[decorate,decoration={snake,amplitude=.4mm,segment length=2mm}] (X) -- node[left,font=\small] {CA} (mid.south west);
    \end{tikzpicture} 
  };
\end{diagram}
\end{widetext}
where ``CA'' is short for ``Concentric Atomic,'' indicating that the connected indices are centered on the same atom, and the squiggled lines are used to introduce a shorthand for the left-hand side of the equals sign.  By plugging the robust CADF approximation to ERI, Eq.~\eqref{eq:robustg}, into Eq.~\eqref{eq:K4c2} and relabeling some indices, we can write \exK as
\begin{align}
  K_{\mu_{a}\nu_{c}} &= \tilde K_{\mu_a\nu_c} + \tilde K_{\nu_c\mu_a}, \\
  \tilde K_{\mu_{a}\nu_{c}} &= 
  \begin{multlined}[t]
  \sum_{\si_dX_{(cd)}}\sum_{\la_b}\left[\gbk{\mu_a\la_b}{X_{(cd)}} 
  % phantom for the right paren
  \vphantom{%
  - \tfrac12 \sum_{Y_{(ab)}}C_{\mu_a\la_b}^{Y_{(ab)}}\gbk{Y_{(ab)}}{X_{(cd)}}
  }\right.
  \\ 
  % phantom for the left paren
  \left.
  - \tfrac12 \sum_{Y_{(ab)}}C_{\mu_a\la_b}^{Y_{(ab)}}\gbk{Y_{(ab)}}{X_{(cd)}}\right]D^{\la_b\si_d} C_{\nu_c\si_d}^{X_{(cd)}}.
\end{multlined}
\end{align}
Diagrammatically, this can be written
\begin{widetext}
\begin{diagram}[node distance=4mm]{$\tilde K_{\mu_a\nu_c}\,\,:$}[font=\large]
  \node (lpic)  {
    \begin{tikzpicture}[node distance=1cm]
      \node (nu)  { $\nu_c$ };
      \node (si) [left=of nu] { $\si_d$ };
      \node (mid3) at ($ (nu) !0.5! (si) $) {};
      \node (X) [below=of mid3.south] { $X_{(cd)}$ };
      \node (la) [left=of si] { $\la_b$ };
      \node (mu) [left=of la] { $\mu_a$ };
      \draw (mu) -- node[above,font=\small] (S) {S} (la);
      \draw (si) -- node[above,font=\small] (P) {P} (la);
      \draw (si) -- node[above,font=\small] (S) {S} (nu);
      \draw[decorate,decoration={snake,amplitude=.4mm,segment length=2mm}] (X) -- node[left,font=\small] {CA} (mid3.south);
    \end{tikzpicture} 
  };
  \node (minus) [font=\Large,right=of lpic] {$-\ \ \frac12$};
  \node (rpic) [right=of minus] {
    \begin{tikzpicture}[node distance=1cm]
      \node (nu)  { $\nu_c$ };
      \node (si) [left=of nu] { $\si_d$ };
      \node (mid) at ($ (nu) !0.5! (si) $) {};
      \node (X) [below=of mid.south west] { $X_{(cd)}$ };
      \node (la) [left=of si] { $\la_b$ };
      \node (mu) [left=of la] { $\mu_a$ };
      \node (mid2) at ($ (mu) !0.5! (la) $) {};
      \node (Y) [below=of mid2.south west] { $Y_{(ab)}$ };
      \draw (mu) -- node[above,font=\small] (S) {S} (la);
      \draw (si) -- node[above,font=\small] (P) {P} (la);
      \draw (si) -- node[above,font=\small] (S) {S} (nu);
      \draw[decorate,decoration={snake,amplitude=.4mm,segment length=2mm}] (Y) -- node[left,font=\small] {CA} (mid2.south west);
      \draw[decorate,decoration={snake,amplitude=.4mm,segment length=2mm}] (X) -- node[left,font=\small] {CA} (mid.south west);
    \end{tikzpicture} 
  };
\end{diagram}
\end{widetext}
Both terms in the diagram are connected, hence a linear-scaling algorithm can be designed. However, it is possible to significantly reduce the computational cost of the first term by screening the three-center two-electron integrals.

\subsubsection{Three-Center Integral Screening with SQV$\ell$}

Three-center ERIs can be screened more efficiently than the standard four-center ERIs because the
potential created by a solid-harmonic Gaussian $|X)$ of angular momentum $l_{X}$ decays as $R^{-(l_{X}+1)}$, rather than the $R^{-1}$ decay of a general product $|\nu\si)$. The decay of the Coulomb operator does not matter as much for the four-center ERI-based exchange construction because the density decays so rapidly. In the CADF-based approach, efficient screening of the three-center ERIs is important and cannot be achieved using only the Schwartz bound. Recently we developed a tight estimator, called SQV$\ell$, for three-center Coulomb integrals that takes into account the correct asymptotic decay of the potential:\footnote{D.~S.~Hollman, H.~F.~Schaefer, and E.~F.~Valeev, J.~Chem.~Phys.~(submitted)}
\begin{align}
  \gbk{\kappa\eta}{\theta}\approx 
  \begin{dcases}
    %\frac{S_{ab}\,\pi\sqrt{2(2\ell_c-1)!!}}{R_{Ac}^{\ell_c+1}\zeta_c^{\frac{2\ell_c+3}{4}}(\zeta_a + \zeta_b)^{\frac14}}, & 
    (2\pi)^{3/4}\,\beta_{\ell_\theta}\!(\zeta_\theta)\frac{|S_{\kappa\eta}|}{R^{\ell_\theta+1}}
    & \begin{multlined}[c]
      R > \ext_{\kappa\eta} + \ext_{\theta}  \\ \text{and } S_{\kappa\eta}/Q_{\kappa\eta} > \thsq 
    \end{multlined} 
    \\
    %\frac{Q_{ab}\,\pi\sqrt{2(2\ell_c-1)!!}}{R_{Ac}^{\ell_c+1}\zeta_c^{\frac{2\ell_c+3}{4}}(\zeta_a + \zeta_b)^{\frac14}}, 
    \frac{\pi\sqrt2\,\beta_{\ell_\theta}\!(\zeta_\theta)\,Q_{\kappa\eta}}{(\zeta_\kappa+\zeta_\eta)^{\frac14}R^{\ell_\theta+1}}
    & \begin{multlined}[c]
      R > \ext_{\kappa\eta} + \ext_{\theta} \\ \text{and } S_{\kappa\eta}/Q_{\kappa\eta} \leq \thsq 
    \end{multlined} 
    \\
    \ Q_{\kappa\eta}Q_\theta &  R \leq \ext_{\kappa\eta} + \ext_{\theta},
  \end{dcases}
  \label{eq:sqvl}
\end{align}
where 
\begin{align}
  \beta_{\ell}(\zeta) &\equiv \zeta^{-\frac{2\ell+3}4}\sqrt{(2\ell-1)!!}, \\
  Q_{\mu\nu} &\equiv \sbk{\mu\nu},
\end{align}
$\zeta_\kappa$, $\zeta_\eta$, and $\zeta_\theta$ are the exponents of the various primitives, $S_{\kappa\eta}$ is the overlap, $\ext_{\kappa\eta}$ are CFMM extents, and $\thws$ and $\thsq$ are user-defined thresholds (see Ref.~\citenum{Note1} for details, including generalization to contracted basis functions).  Practically speaking, the ratio of prefactors from the first two cases is folded into the $S_{\kappa\eta}/Q_{\kappa\eta}$ ratio, and the second case is used if this ratio is less than one.  Effectively, the far field estimator uses the minimum of the first two cases when the $S_{\kappa\eta}/Q_{\kappa\eta}$ ratio is greater than \thsq, and only the second case when the ratio is less than \thsq.  This latter detail will be important for the use of this estimator in Section~\ref{sec:link}.

The SQV$\ell$ estimator allows us to add a crucial edge to the CADF exchange screening diagram:
\begin{diagram}{$\tilde K_{\mu_a\nu_c}\,\,\leftarrow:$}
  \node (nu)  { $\nu_c$ };
  \node (si) [left=of nu] { $\si_d$ };
  \node (mid) at ($ (nu) !0.5! (si) $) {};
  \node (X) [below=of mid.south] { $X_{(cd)}$ };
  \node (la) [left=of si] { $\la_b$ };
  \node (mu) [left=of la] { $\mu_a$ };
  \node (mid2) at ($ (mu) !0.5! (la) $) {};
  \draw (mu) -- node[above,font=\small] (S) {S} (la);
  \draw (si) -- node[above,font=\small] (P) {P} (la);
  \draw (si) -- node[above,font=\small] (S) {S} (nu);
  \draw[decorate,decoration={snake,amplitude=.4mm,segment length=2mm}] (X) -- node[left,font=\small] {CA} (mid.south);
  \draw[dashed] (X) .. controls (X-|la) and (X-|mid2.south) .. node [below,near start,font=\small] {$R^{-1-\ell_X}$} (mid2.south);
  \label{d:sqvl}
\end{diagram}
\hspace{-0.8em} Note that the efficiency of the SQV$\ell$ screening will increase for kets with higher angular momenta; hence, the more expensive the three-center ERI, the more likely it will be screened out.
%Unlike the original QQR estimator, this new estimator has a discontinuity at $R = \ext'_{\mu\nu} + \ext'_{X}$, the implications of which are discussed in Appendix \ref{ap:discont}.

%%%%%%%%%%%%%%%%%%%%%%%%%%%%%%%%%%%%%%%%%%%%%%%%%%%%%%%%%%%%%%%%%%%%%%%%%%%%%%%%%%%%%%%%%%%%%%%%%%%%%%%%%%%%%%%%%%%%%%%%%%%%%%%%%%%
%%%%%%%%%%%%%%%%%%%%%%%%%%%%%%%%%%%%%%%%%%%%%%%%%%%%%%%%%%%%%%%%%%%%%%%%%%%%%%%%%%%%%%%%%%%%%%%%%%%%%%%%%%%%%%%%%%%%%%%%%%%%%%%%%%%
%%%%%%%%%%%%%%%%%%%%%%%%%%%%%%%%%%%%%%%%%%%%%%%%%%%%%%%%%%%%%%%%%%%%%%%%%%%%%%%%%%%%%%%%%%%%%%%%%%%%%%%%%%%%%%%%%%%%%%%%%%%%%%%%%%%

\section{\bigOhNone ~CADF Exchange Algorithm}
\label{sec:link}

Like four-center ERI-based linear scaling exchange algorithms,\cite{Ochsenfeld:1998p1663,Schwegler:1997p9708} our linear scaling exchange algorithm based on concentric atomic density fitting (which we will call CADF-LinK) relies on the construction of prescreening lists, $\Lst3{\mu X}$ and $\Lst B{\mu X}$, to drive the exchange matrix build.  Both sets of lists are computed with linear effort.

%%%%%%%%%%%%%%%%%%%%%%%%%%%%%%%%%%%%%%%%%%%%%%%%%%%%%%%%%%%%%%%%%%%%%%%%%%%%%%%%%%%%%%%%%%%%%%%%%%%%%%%%%%%%%%%%%%%%%%%%%%%%%%%%%%%
%%%%%%%%%%%%%%%%%%%%%%%%%%%%%%%%%%%%%%%%%%%%%%%%%%%%%%%%%%%%%%%%%%%%%%%%%%%%%%%%%%%%%%%%%%%%%%%%%%%%%%%%%%%%%%%%%%%%%%%%%%%%%%%%%%%

\newcommand{\distfact}[3]{\ensuremath{\tilde R_{#1#2}^{#3}}\xspace}
\newcommand{\distf}{\distfact{\mu}{\la}{X}}
\subsection{The $\Lst3{\mu X}$ Lists}
$\Lst3{\mu X}$ specifies the list of $\la$ for a given $(\mu X)$ pair for which $\gbk{\mu\la}{X}$ must be computed.  
%\edsays{Are $\mu$ and $X$ shells or functions? If former, what does $(X|X)^{1/2}$ mean, etc.? }
It is defined as follows:
\begin{align}
  \la \in \Lst3{\mu X} \ \Leftrightarrow\ \dbar_{\la}^X \distf > \epsilon,
\end{align}
where
\begin{align}
  \dbar_{\la}^X &\equiv \sum_\sigma^{OBS} |D^{\la\sigma}| \bar C_\sigma^X,
  \label{eq:dbar} \\
  \bar C_{\sigma_d}^{X_c} &\equiv \gbk XX^{1/2}\frob{\nu_e\in OBS}{C_{\nu_e\sigma_d}^{X_c}\delta_{c\in(de)}} \\
  &\begin{multlined}
    = \gbk XX^{1/2}\left[\frob{\nu_c \in (c)_{OBS}}{C_{\nu_c\sigma_d}^{X_c}} \right.
    \\ \left. + \frob{\nu_e\in OBS}{C_{\nu_e\sigma_d}^{X_c}}\delta_{cd}
     - \frob{\nu_c \in (c)_{OBS}}{C_{\nu_c\sigma_d}^{X_c}}\delta_{cd}\right],
    \label{eq:cexpand}
  \end{multlined} \\
  \distf &\equiv \gbk XX^{-1/2} I_{\mathrm{\sqvl}}(\mu,\la,X) 
\end{align}
and $I_{\mathrm{\sqvl}}(\mu,\la,X)$ is the \sqvl screening factor from Eq.~\eqref{eq:sqvl}.  (Keep in mind that though the tensor-like notation \distf is used for compactness, this quantity is intended to be computed on the fly, so a notation like $\tilde R(\mu,\la,X)$ would be conceptually more appropriate.)   We have loosely used the notation 
\begin{align}
  \frob{x\in \mathbb S}{f_x} = \sqrt{\sum_{x\in \mathbb S}{f_x^2}}
\end{align}
to represent the block Frobenius norm of a matrix or tensor expression.  In these expressions and the following, the orbital basis set (OBS) and density fitting basis set (DFBS) are distinguished by their respective abbreviations, and the notation $(c)_{OBS}$, for instance, represents the set of all basis functions from the OBS centered on atom  $c$.  The density fitting coefficients $C_{\nu_e\sigma_d}^{X_c}$ can be obtained in linear effort after a Schwarz-based prescreening of OBS pairs, as discussed in the original CADF paper.\cite{Hollman:2014gc}

We immediately note that (theoretically, at least) the $\ \mathbf\dbar$ (``d-bar'') intermediate can be constructed in linear scaling time, since
\begin{diagram}{$\dbar_{\la}^{X}\,\,:$}
  \node (nu)  { $\nu_c$ };
  \node (si) [left=of nu] { $\si_d$ };
  \node (mid) at ($ (nu) !0.5! (si) $) {};
  \node (X) [below=of mid.south] { $X_{(cd)}$ };
  \node (la) [left=of si] { $\la_b$ };
  \draw (si) -- node[above,font=\small] (P) {P} (la);
  \draw (si) -- node[above,font=\small] (S) {S} (nu);
  \draw[decorate,decoration={snake,amplitude=.4mm,segment length=2mm}] (X) -- node[left,font=\small] {CA} (mid.south);
\end{diagram}
\hspace{-0.8em} is a fully connected diagram.  Indeed, the construction of $\ \mathbf\dbar$ is quite straightforward if the density matrix \Dmat is stored in a sparse data structure.  The construction of $\mathbf{\bar C}$ can also be performed in linear scaling effort, since the norm in the second term of Eq.~\eqref{eq:cexpand} need only loop over the $\nu_e$ that have significant overlap with $\si_d$.  With this in mind, a procedure for the linear scaling construction of $\Lstset3$ is given in Algorithm~\ref{alg:l3}.  Similar to the original \LinK algorithm,\cite{Ochsenfeld:1998p1663} the key to linear scaling in Algorithm~\ref{alg:l3} is the list ordering that allows the program to exit the loops early in lines~\ref{al:exit1} and \ref{al:exit2}.  The auxiliary list set $\Lst{DC}{\la}$ is composed of $X$ for which $\dbar_\la^X$ is significant; each list is then sorted by decreasing $\dbar_\la^X$ value.  The other helper list set, $\Lst{S}\la$, is simply the $\mu$ that form Schwarz pairs with $\la$, sorted by decreasing value of $|\gbk{\mu\la}{\mu\la}|^{1/2}$.

\begin{algorithm}[t]
  \caption{Build \Lstset3}%
  \label{alg:l3}
  \begin{algorithmic}[1]
    \ForIn[$\la$]{$OBS$} 
      \myState $\Lst{DC}{\la} \leftarrow$ $X\in DFBS$ with $\dbar_\la^X>\epsilon_d$ \;
      \myState Sort $\Lst{DC}{\la}$ by decreasing $|\dbar_\la^X|$ \;
      \ForIn[$X$]{$\Lst{DC}{\la}$}
        \ForIn[$\mu$]{$\Lst{S}{\la}$}
          \If{$\dbar_\la^X\gbk{\mu\la}{\mu\la}^{1/2} > \epsilon_K$}
            \myState Compute distance factor \distf \;
            \If{$\dbar_\la^X \distf > \epsilon_K$}
              \myState Add $\la$ to $\Lst3{\mu X}$ \;
            \EndIf
          \Else
            \myState \textbf{break} out of $\mu$ loop \label{al:exit1} \;
          \EndIf
        \EndForIn{$\mu$}
        \If{$\mu$ loop exited on first iteration}
          \myState \textbf{break} out of $X$ loop \label{al:exit2} \;
        \EndIf
      \EndForIn{$X$}
    \EndForIn{$\la$}
  \end{algorithmic}
\end{algorithm}

\subsubsection{Quadratic Exchange Matrix Build}

Using only the \Lstset3 lists, a quadratic scaling \exK build algorithm can be devised.  Such a procedure is outlined in Algorithm~\ref{alg:krepl}.  Note that the subscripted indices (e.g., $\mu_a$) are descriptive rather than prescriptive:  the subscript $a$ merely indicates that the center on which the function $\mu$ is located has the index $a$; it does not indicate anything specific about the iteration range of $\mu_a$.  The algorithm loops over the significant $(\mu \la | X)$ integrals and immediately contracts these contributions into the \exB intermediate.   The quadratic scaling of this algorithm arises because the loop in line~\ref{al:bloop} is not restricted; the restriction of this loop requires $\Lstset B$, discussed in Section \ref{sec:lstb}.

\begin{algorithm}[t]
  \caption{Quadratic \exK build using \Lstset3}%
  \label{alg:krepl}
  \begin{algorithmic}[1]
    \myState Form \Lst3{\mu X} for significant $(\mu_a X_c)$ \Comment{See Algorithm \ref{alg:l3}} \;
    \ForWith[$(\mu_a X_c)$]{$\Lst{3}{\mu X} \ne \varnothing$}
      \ForIn[$\la_b$]{\Lst3{\mu X}}
        \myState Compute $\gbk{\mu_a\la_b}{X_c}$ \;
        \myState $\bar g_{\mu_a\la_b}^{X_c} = \gbk{\mu_a\la_b}{X_c}$ \;
        \ForIn[$Y_{(ab)}$]{$(a)_{DFBS} \cup (b)_{DFBS}$}
          \myState $\bar g_{\mu_a\la_b}^{X_c} \meq \frac12 C_{\mu_a\la_b}^{Y_{(ab)}} \gbk{Y_{(ab)}}{X_c}$ \; \vspace{0.6ex}
        \EndForIn{$Y$}
        \ForIn[$\si_d$]{OBS}
          \myState $B_{\mu_a\sigma_d}^{X_c} \peq \bar{g}_{\mu_a\la_b}^{X_c}D^{\la_b\sigma_d}$ \label{al:bloop} \; \vspace{0.6ex}
        \EndForIn{$\si_d$}
      \EndForIn{$\la_b$}
      \ForIn[$\nu_c$]{$(c)_{OBS}$} \label{al:kbegin}
        \ForIn[$\si_d$]{$\Lst S{\nu}$ \textbf{with} $\sigma_d \notin (c)_{OBS}$ }
          \myState$\tilde K_{\mu_a\nu_c} \peq C_{\nu_c\si_d}^{X_c} B_{\mu_a\si_d}^{X_c}$\label{al:kpart1} \; \vspace{0.6ex}
        \EndForIn{$\si_d$}
      \EndForIn{$\nu_c$}
      \ForIn[$\si_c$]{$(c)_{OBS}$}
        \ForIn[$\nu_d$]{$\Lst{S}{\si}$}
          \myState $\tilde K_{\mu_a\nu_d} \peq C_{\si_c\nu_d}^{X_c} B_{\mu_a\si_c}^{X_c}$\label{al:kpart2} \; \vspace{0.6ex}
        \EndForIn{$\nu_d$}
      \EndForIn{$\si_c$} \label{al:kend}
    \EndForIn{$(\mu_a X_c)$}
    \myState ${\exK} = \tilde{\exK} + \tilde{\exK}^\top$ \;
  \end{algorithmic}
\end{algorithm}

Several important aspects of this quadratic scaling algorithm merit further discussion before the introduction of the linear scaling version.  First, note that the summation of $\exB$ into $\tilde \exK$ (lines \ref{al:kbegin}--\ref{al:kend} of Algorithm~\ref{alg:krepl}) is split into two loops with one index restricted in each loop.  This restriction is made possible by the CADF coefficient definition, which precludes coefficients with DFBS functions that do not share a center with the OBS pair being fit.  This loop structure also makes it possible to store the coefficients in an ordering more efficient for vectorization: $C_{\nu_c\si_d}^{X_c}$ is stored as $X_c\rightarrow [\nu_c \text{\ offset in }(c)_{OBS}] \rightarrow \si_d$ (from slowest- to fastest-running index), which is still a $\bigOhN2$ data structure.  Secondly, note that the algorithm is integral direct, and that the algorithm's storage requirements never exceed \bigOhN2.  The intermediate $\exB$ requires linear storage in this algorithm, and the integrals $\gbk{\mu\la}{X}$ need not be stored outside of the inner loop in which they are computed.

%%%%%%%%%%%%%%%%%%%%%%%%%%%%%%%%%%%%%%%%%%%%%%%%%%%%%%%%%%%%%%%%%%%%%%%%%%%%%%%%%%%%%%%%%%%%%%%%%%%%%%%%%%%%%%%%%%%%%%%%%%%%%%%%%%%%%%%%%%%%%%%%%%%%%%%%%%%%%%%%
%%%%%%%%%%%%%%%%%%%%%%%%%%%%%%%%%%%%%%%%%%%%%%%%%%%%%%%%%%%%%%%%%%%%%%%%%%%%%%%%%%%%%%%%%%%%%%%%%%%%%%%%%%%%%%%%%%%%%%%%%%%%%%%%%%%%%%%%%%%%%%%%%%%%%%%%%%%%%%%%

\subsection{The \Lst B{\mu X} Lists}
\label{sec:lstb}

In order to further improve the scaling of Algorithm \ref{alg:krepl} to linear, another set of prescreening lists, $\Lstset B$, must be formed.  For a given $\mu$ and $X$,  the list $\Lst B{\mu X}$ from the set \Lstset B is defined as
\begin{align}
  \si \in \Lst B{\mu X} \Leftrightarrow \bar C_{\si}^{X} \bbar^\si_{(\mu X)} > \epsilon 
\end{align}
where
\begin{align}
  \bbar^\si_{(\mu X)} \equiv \sum_{\la\in\Lst3{\mu X}} |D^{\la\si}|\distf .
\end{align}
As with $\mathbf \dbar$, the formation of $\mathbf \bbar$ in linear effort is relatively trivial if $\Dmat$ is stored in a sparse data structure, and particularly if $\Lst3{\mu X}$ is sorted by decreasing \distf.  An algorithm for the linear scaling formation of \Lstset B is outlined in Algorithm~\ref{alg:lb}.  The procedure is greatly simplified by the fact that a list of significant $(\mu X)$ pairs is already known from the construction of \Lstset 3.  The algorithm is little more than a sparse-sparse matrix multiply.  If $\mathbf{\bar C}$ is stored in a sparse data structure, line~\ref{al:lC} is already part of the manifestation of $\mathbf{\bar C}$ in memory.  Similarly, the conditional in line~\ref{al:lbif} is essentially the procedure that a sparse-sparse matrix multiply undergoes to determine the nontrivial entries in the product matrix.  Nonetheless, we have included the \Lstset B build algorithm here for completeness.

\begin{algorithm}
  \caption{Build \Lstset B}
  \label{alg:lb}
  \begin{algorithmic}[1]
    \ForIn[$X$]{$OBS$}
      \myState $\Lst{\bar C}X \leftarrow \si \in OBS$ with $\bar C_\si^X > \epsilon_{\bar C}$ \label{al:lC} \;
    \EndForIn{$\si$}
    \ForWith[$(\mu X)$]{$\Lst3{\mu X}\ne \varnothing$}
      \ForIn[$\si$]{$\Lst{\bar C}X$}      
      \If{$\bar C_{\si}^X \bbar^\si_{(\mu X)} > \epsilon_K$} \label{al:lbif}
          \myState Add $\si$ to $\Lst B{\mu X}$ \;
        \EndIf
      \EndForIn{$\si$}
    \EndForIn{$(\mu X)$}
  \end{algorithmic}
\end{algorithm}

\subsubsection{Linear Scaling Exchange Matrix Build}

The use of \Lstset B to create a \exK build procedure that scales linearly requires only a few modifications to Algorithm~\ref{alg:krepl}.  The revised procedure is given in Algorithm~\ref{alg:kb}.  The main changes involve restricting the loops in lines \ref{al:blr} and \ref{al:klr}.  The restriction in line \ref{al:blr} amounts to excluding from \exB any $\sigma_d$ where the contraction over the full set of significant $\la_b$ for a given $(\mu_aX_c)$ is negligible.  These indices can then also be excluded from the summation in line \ref{al:klr}, since the $B_{\mu_a\si_d}^{X_c}$ are trivial for these $\si_d$.  Further optimization could be made by restricting the loop in line \ref{al:blr} to include only $\si_d$ for a given $\la_b$ where $D^{\la_b\si_d}$ is significant.  This optimization would arise naturally if a sparse data structure were used for \Dmat, but in the present implementation dense \bigOhN2 data structures were used (see Section \ref{sec:impl}).  

\begin{algorithm}[t]
  \caption{Linear Scaling \exK build using \Lstset3 and \Lstset B}%
  \label{alg:kb}
  \begin{algorithmic}[1]
    \myState Form \Lst3{\mu X} for significant $(\mu_a X_c)$ \Comment{See Algorithm \ref{alg:l3}} \;
    \myState Form \Lst B{\mu X} for significant $(\mu_a X_c)$ \Comment{See Algorithm \ref{alg:lb}} \;
    \ForWith[$(\mu_a X_c)$]{$\Lst{3}{\mu X} \ne \varnothing$ and $\Lst{B}{\mu X} \ne \varnothing$}
      \ForIn[$\la_b$]{\Lst3{\mu X}} \label{al:3lr}
        \myState Compute $\gbk{\mu_a\la_b}{X_c}$ \; \label{al:intc}
        \myState $\bar g_{\mu_a\la_b}^{X_c} = \gbk{\mu_a\la_b}{X_c}$ \;
        \ForIn[$Y_{(ab)}$]{$(a)_{DFBS} \cup (b)_{DFBS}$}
          \myState $\bar g_{\mu_a\la_b}^{X_c} \meq \frac12 C_{\mu_a\la_b}^{Y_{(ab)}} \gbk{Y_{(ab)}}{X_c}$ \; \vspace{0.6ex}
        \EndForIn{$Y$}
        \ForIn[$\si_d$]{\Lst B{\mu X}} \label{al:blr}
        \myState $B_{\mu_a\sigma_d}^{X_c} \peq \bar{g}_{\mu_a\la_b}^{X_c}D^{\la_b\sigma_d}$ \; \label{al:bcon} \vspace{0.6ex}
        \EndForIn{$\si_d$}
      \EndForIn{$\la_b$}
      \ForIn[$\nu_c$]{$(c)_{OBS}$} 
        \ForIn[$\si_d$]{$\Lst B{\mu X}$ \textbf{with} $\sigma_d \notin (c)_{OBS}$ } \label{al:klr}
        \myState$\tilde K_{\mu_a\nu_c} \peq C_{\nu_c\si_d}^{X_c} B_{\mu_a\si_d}^{X_c}$ \; \label{al:kcon1} \vspace{0.6ex}
        \EndForIn{$\si_d$}
      \EndForIn{$\nu_c$}
      \ForIn[$\si_c$]{$(c)_{OBS}\cap \Lst B{\mu X}$} \label{al:k2out}
      \ForIn[$\nu_d$]{$\Lst{S}{\si}$} \label{al:schwloop}
          \myState $\tilde K_{\mu_a\nu_d} \peq C_{\si_c\nu_d}^{X_c} B_{\mu_a\si_c}^{X_c}$ \; \label{al:kcon2} \vspace{0.6ex}
        \EndForIn{$\nu_d$}
      \EndForIn{$\si_c$} 
    \EndForIn{$(\mu_a X_c)$}
    \myState ${\exK} = \tilde{\exK} + \tilde{\exK}^\top$ \;
  \end{algorithmic}
\end{algorithm}

%%%%%%%%%%%%%%%%%%%%%%%%%%%%%%%%%%%%%%%%%%%%%%%%%%%%%%%%%%%%%%%%%%%%%%%%%%%%%%%%
%%%%%%%%%%%%%%%%%%%%%%%%%%%%%%%%%%%%%%%%%%%%%%%%%%%%%%%%%%%%%%%%%%%%%%%%%%%%%%%%
%%%%%%%%%%%%%%%%%%%%%%%%%%%%%%%%%%%%%%%%%%%%%%%%%%%%%%%%%%%%%%%%%%%%%%%%%%%%%%%%
%%%%%%%%%%%%%%%%%%%%%%%%%%%%%%%%%%%%%%%%%%%%%%%%%%%%%%%%%%%%%%%%%%%%%%%%%%%%%%%%
%%%%%%%%%%%%%%%%%%%%%%%%%%%%%%%%%%%%%%%%%%%%%%%%%%%%%%%%%%%%%%%%%%%%%%%%%%%%%%%%
%%%%%%%%%%%%%%%%%%%%%%%%%%%%%%%%%%%%%%%%%%%%%%%%%%%%%%%%%%%%%%%%%%%%%%%%%%%%%%%%
%%%%%%%%%%%%%%%%%%%%%%%%%%%%%%%%%%%%%%%%%%%%%%%%%%%%%%%%%%%%%%%%%%%%%%%%%%%%%%%%

\section{Computational Details}
\label{sec:impl}

The algorithms detailed in Section~\ref{sec:link} were implemented in a development version (commit tag {\tt 3.0.0-cadflink} of the {\tt localdf} branch) of the Massively Parallel Quantum Chemistry ({\tt MPQC})\cite{mpqc} quantum chemistry package.
While the algorithms were implemented predominantly as written here, several small details differ from the ideal linear scaling implementation.  Most prominently, since integrals are computed more efficiently in shell blocks, all of the indices in the algorithms actually represent shell blocks rather than individual basis functions.   In other words, for instance,
\begin{align}
  |\gbk XX|^{1/2} = \frob{X'\in X}{|\gbk {X'}{X'}|^{1/2}}
\end{align}
where $X'$ is a function index and $X$ is a shell index.
The upshot of this is that the prescreening algorithms and list formations are significantly less expensive than the main computation in practice, even though in principle the onset of linear scaling is later for these portions of the computation.  (The prescreening is more expensive in principle because the exit conditions of the loops in Algorithm~\ref{alg:l3} use the Schwarz estimate of the integrals rather than the distance-including estimate, since the latter cannot be easily ordered).  The quantities used in the screening process are replaced by their shell block Frobenius norm analogs.  Indeed, if a scenario were to arise in which the screening portions of the algorithm began to dominate the cost, the concept of shell blocks could be generalized further to arbitrary blocks and a multi-tiered prescreening approach could be used.   Another difference in our implementation is that dense data structures were used to store all \bigOhN2 quantities.  The primary reason for this is that the SCF solver in {\tt MPQC} is currently based on an \bigOhN3 eigensolve; alternative \bigOhNone ~SCF solvers are well known (e.g.,~density matrix minimization\cite{Challacombe:1999p2332}) but are not yet implemented in our program.

It is difficult to measure the practical scaling of an algorithm outside of the context of its implementation details and execution environment.
Thus, many authors\cite{Beer:2008p221102,Maurer:2012p144107,Maurer:2013p014101,Scuseria:1999p8330} opt to present their algorithmic scaling in terms of integral counts, contraction sizes, or other implementation- and execution-independent metrics.  Here we use several such metrics
corresponding to particular sections of Algorithm \ref{alg:kb}. Our implementation of CADF-LinK has been heavily optimized to run well on massively parallel computers and to take advantage of both thread and process parallelism.  However, a thorough discussion of the challenges involved in the parallelization of CADF-LinK has been reserved for a separate paper\footnote{D.~S.~Hollman, H.~F.~Schaefer, and E. F. Valeev, in preparation} in the interest of both saving space and appealing to a more general audience.
% TODO decide whether to remove reference to future work

For our cost metrics, we chose a series of one dimensional molecules (linear alkanes) and a series of three dimensional systems (water clusters).  Cartesian coordinates are included in the supplemental information.  We used the basis set pairs Def2-SVP\cite{Weigend:2005dh}/Def2-SVP/JK,\cite{Weigend:2006kv} cc-pVTZ\cite{Dunning:1989p1007}/cc-pVTZ/JK,\cite{Weigend:2002p4285} and cc-pVQZ/cc-pVQZ/JK.  For these series of molecules and basis sets, we recorded the number of three-center integrals computed (line \ref{al:intc}), the number of multiplies in the contraction to form $\exB$ (line \ref{al:bcon}), and the number of multiplies in the contractions to form $\tilde\exK$ (lines \ref{al:kcon1} and \ref{al:kcon2}).   These data were averaged over the first three SCF iterations; later iterations were omitted to minimize variability due to convergence acceleration procedures.  

For the computations in the remainder of this work, an SCF convergence criterion of $10^{-6}$ was used.  Our method has no difficulty converging further than this, but since the emphasis is on per iteration performance, we chose a relatively loose convergence threshold.  An initial screening threshold $\epsilon_K$ of $10^{-6}$ was used throughout as well.  However, this number was varied for differential density iterations as follows:  the screening threshold for a given iteration was taken to be the initial threshold if the full density was used or the ratio of differential density to full density Frobenius norm times the initial threshold if a differential density was used, down to a minimum threshold of $10^{-11}$.  For the \sqvl estimate, $10^{-1}$ was used for both $\thws$ and $\thsq$.\cite{Note1}  This use of a scaled screening threshold allowed us to use a quite aggressive initial threshold without sacrificing as much accuracy in the final result.  This amounts to a simplified version of a previous variable precision SCF approach\cite{Luehr:2011jr} that yields excellent results for large molecules.  Indeed, we were usually able to converge SCF computations to a root mean squared density change of $10^{-10}$ or less using an initial threshold of $10^{-6}$.  This thresholding scheme also spreads the workload across iterations in a fairly uniform manor:  in almost all cases, no iteration costs more than a factor of 2 different from any other iteration (without threshold scaling, factors of 5-10 were often observed).

%%%%%%%%%%%%%%%%%%%%%%%%%%%%%%%%%%%%%%%%%%%%%%%%%%%%%%%%%%%%%%%%%%%%%%%%%%%%%%%%
%%%%%%%%%%%%%%%%%%%%%%%%%%%%%%%%%%%%%%%%%%%%%%%%%%%%%%%%%%%%%%%%%%%%%%%%%%%%%%%%
%%%%%%%%%%%%%%%%%%%%%%%%%%%%%%%%%%%%%%%%%%%%%%%%%%%%%%%%%%%%%%%%%%%%%%%%%%%%%%%%
%%%%%%%%%%%%%%%%%%%%%%%%%%%%%%%%%%%%%%%%%%%%%%%%%%%%%%%%%%%%%%%%%%%%%%%%%%%%%%%%
%%%%%%%%%%%%%%%%%%%%%%%%%%%%%%%%%%%%%%%%%%%%%%%%%%%%%%%%%%%%%%%%%%%%%%%%%%%%%%%%
%%%%%%%%%%%%%%%%%%%%%%%%%%%%%%%%%%%%%%%%%%%%%%%%%%%%%%%%%%%%%%%%%%%%%%%%%%%%%%%%
%%%%%%%%%%%%%%%%%%%%%%%%%%%%%%%%%%%%%%%%%%%%%%%%%%%%%%%%%%%%%%%%%%%%%%%%%%%%%%%%

\section{Results and Discussion}
\label{sec:results}

\subsection{Scaling with respect to cost metrics}

\begin{figure}
  \centering
  \includegraphics[width=0.5\textwidth]{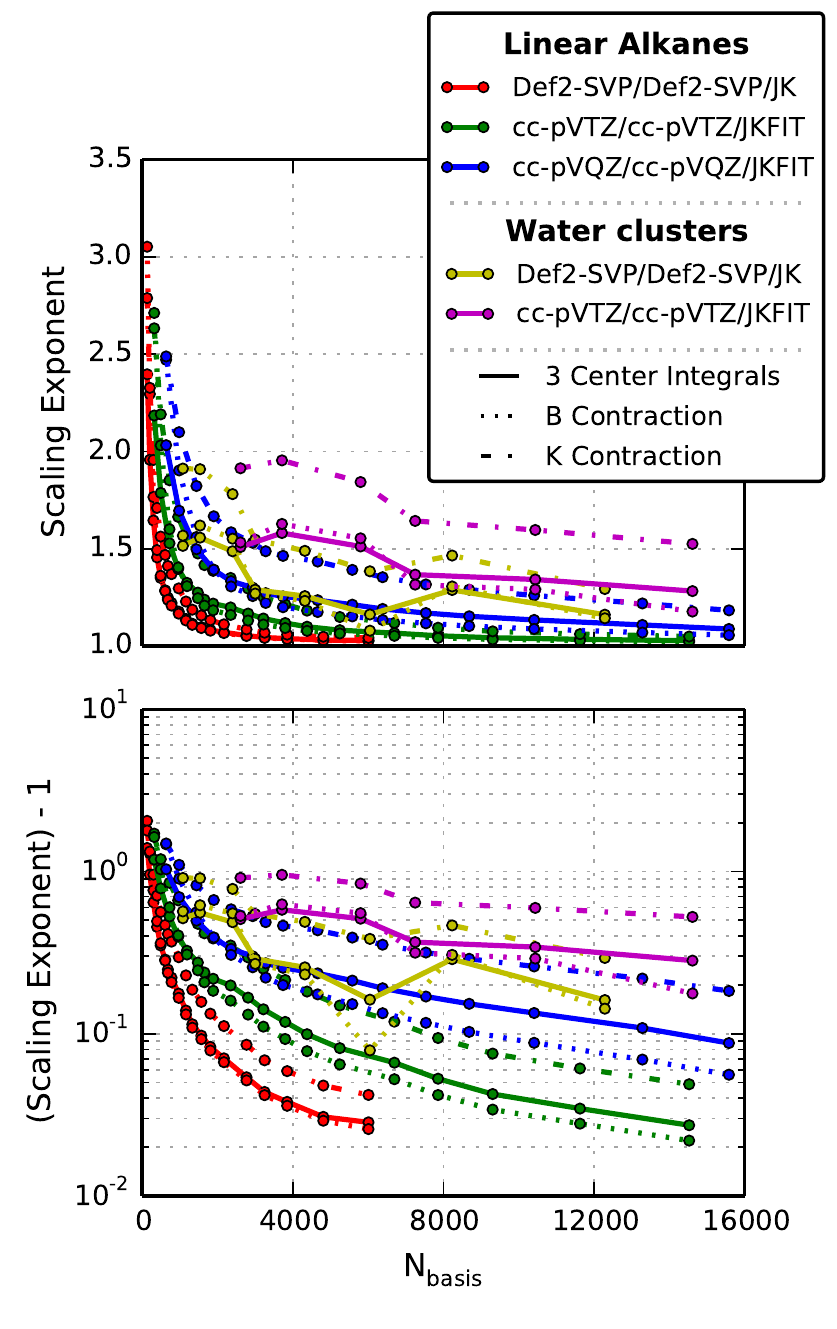}
  \caption{Effective scaling exponents (Eq. \eqref{eq:kNi}) of the three cost metrics of the CADF-LinK algorithm described in Section~\ref{sec:impl}.  The top plot uses a linear range, and the bottom plot shows the same data on a logarithmic range (subtracting 1 to show proximity to linear scaling). }
  \label{fig:scaling}
\end{figure}

To analyze the asymptotic scaling of a positive computational cost metric $C(N)$ with the system size parameter $N$ it is convenient to
introduce of an effective scaling exponent:
\begin{align}
\label{eq:kNi}
k(N_{i}) \approx & \log_{\frac{N_i}{N_{i-1}}}\frac{C(N_i)}{C(N_{i-1})}.
\end{align}
This amounts roughly to a ``two-point fit'' to the form $C(N) = a N^{k} + b$, for constants $a$ and $b$, at the points $N_i$ and $N_{i-1}$.  By definition, the effective exponent of the cost of a \bigOhNone algorithm should approach 1.0 in the limit of large $N$. 

\figref{scaling} shows the effective scaling exponents for the key cost metrics of our algorithm when applied to our one- and three-dimensional series of molecules. It is clear that all scaling exponents are less than 2 for $N>2000$, and decrease monotonically with $N$.
Also, the \exK contraction step has the worst scaling (highest effective exponent). This is anticipated due to the relatively weak restriction of the loop in line~\ref{al:schwloop} of \Algref{kb}, which only utilizes Schwarz screening.  However, since the non-LinK behavior of the $\exK$ contraction is better than other parts of the algorithm (\bigOhN 3 with no screening compared to \bigOhN 4 for the $\exB$ contraction without screening), the \exK contraction still costs less in terms of CPU time than the \exB contraction, even for our largest computations.
Nevertheless, the scaling with respect to the \exK contraction metric is good, reaching \bigOhN{1.1} around 2500 basis functions for linear alkanes with a small basis and \bigOhN{1.2} around 4000 and 10000 basis functions for a triple zeta and quadruple zeta basis, respectively.  Even for three dimensional systems, the scaling of the \exK contraction is still around \bigOhN{1.5} by about 4000 and 15000 basis functions for the Def2-SVP/Def2-SVP/JK and cc-pVTZ/cc-pVTZ/JK basis sets, respectively.  The scaling behavior of the more expensive sections, the integral computation and \exB contraction, is even better.  The scaling reaches \bigOhN{1.1} around 1300, 3500, and 8700 basis functions, respectively, for linear alkanes with the three basis set pairs in our study.  For our three dimensional systems, the scaling of the \exB contraction is about \bigOhN{1.5} by around 2400 and 5800 basis functions for Def2-SVP/Def2-SVP/JK and cc-pVTZ/cc-pVTZ/JK, respectively.

While the scaling exponent functions are less smooth for water clusters compared to linear alkanes due to the less systematic growth of the former, the overall trends in the data are similar for both one- and three-dimensional systems.  Obviously, the decay of the scaling exponents is much more rapid for one dimensional systems and for smaller, less diffuse basis sets.  In the worst case---three dimensional water clusters with the cc-pVTZ/cc-pVTZ/JK basis set---the scaling of the most computationally intense sections of the algorithm is around \bigOhN{1.25} for a system with 756 atoms.  Given that in the context of HF and hybrid KS DFT the typical basis sets are smaller and less diffuse than cc-pVTZ, it completely reasonable to conclude that CADF-LinK will closely approach linear scaling behavior in the vast majority of use cases.

\subsection{Errors}
\label{sec:err}

\begin{figure}
  \centering
  \includegraphics[width=0.4\textwidth]{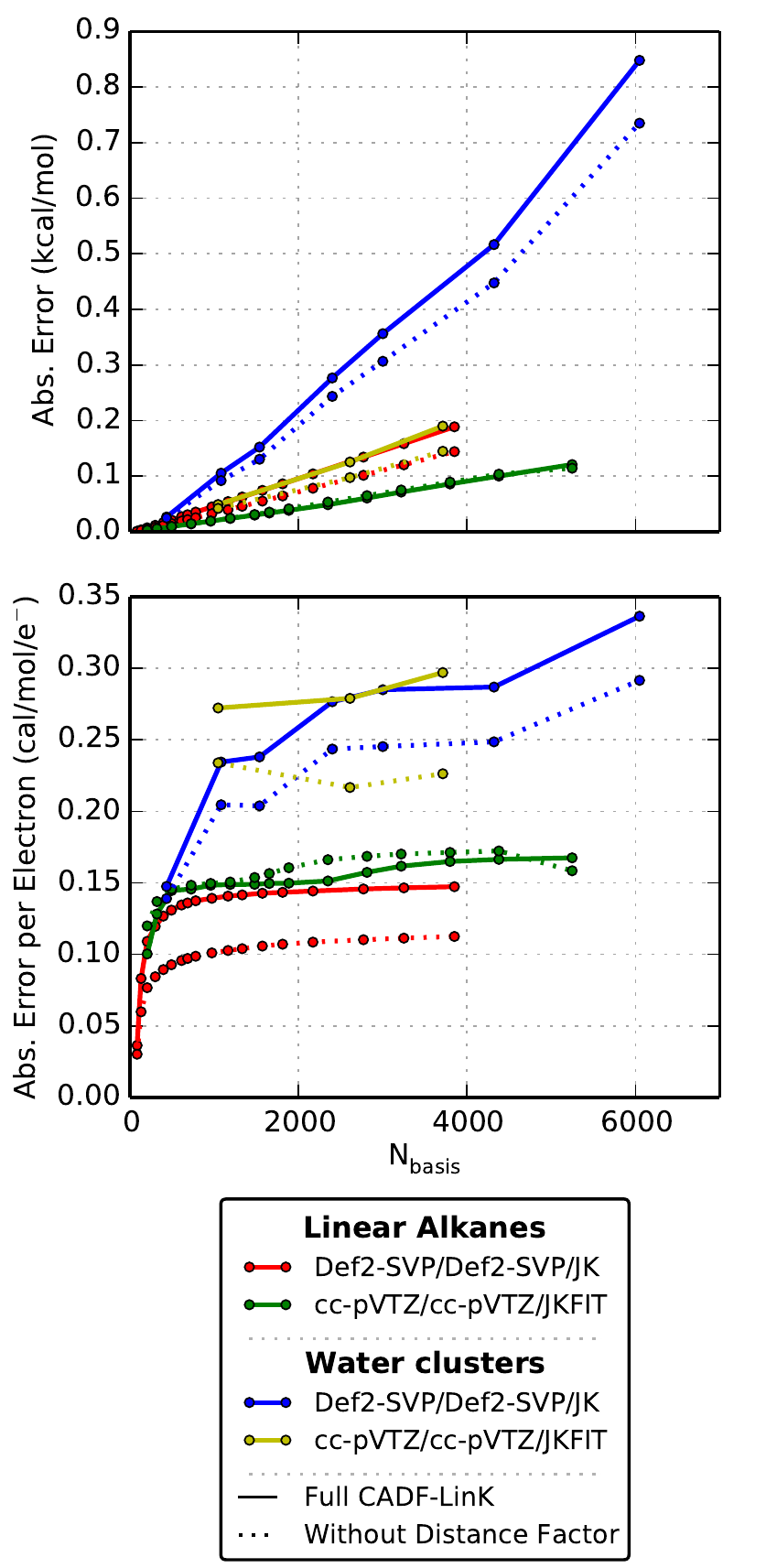}
  \caption{Errors in absolute energies arising from the CADF-LinK approximation, relative to CADF-based exchange using only Schwarz screening.  Errors per electron are given in the bottom plot to show that the CADF-LinK errors increase linearly with system size.}
  \label{fig:errors}
\end{figure}

\figref{errors} shows the errors in absolute energies resulting from the CADF-LinK approximation, relative to CADF-SCF with Schwarz screening only (the Coulomb matrix was computed in the same manner in both sets of computations).  As with the CADF approximation and Hartree--Fock theory itself, the absolute energy errors arising from the CADF-LinK approximation scale linearly with the size of the system.  However, these errors are still about one or two orders of magnitude smaller that the errors arising from the CADF approximation for molecules of comparable size (see Ref.~\citenum{Hollman:2014gc}), which are in turn smaller than the basis set errors or the errors of the Hartree--Fock method itself.  Importantly, as demonstrated by the bottom plot in \figref{errors}, the absolute error per electron does not increase significantly beyond a point for the molecules we tested.

The larger error per electron for three dimensional structures compared to one dimensional ones can be heuristically rationalized as follows.  As two basis functions are separated from each other, their pair-wise contribution (roughly speaking) to the energy is excluded at some distance.  A three dimensional structure will have more basis functions at or near this distance than a one dimensional structure.  Furthermore, in the specific case of water clusters compared to linear alkanes, the latter have a significantly larger number of covalent bonds, meaning that a much smaller portion of the contributions to the energy are likely to have separations near this critical exclusion length.

Another notable feature of \figref{errors} is the error behavior when the distance factor (from the \sqvl estimator) is omitted.  As the original \sqvl paper noted,\cite{Note1} the estimator performs worse for the Def2-$X$VP/Def2-$X$VP/JK basis sets than the cc-pV$X$Z/cc-pV$X$Z/JK basis sets, because the former contains more contracted basis functions, particularly in the fitting basis, than the latter.  Since the \sqvl estimator is tightest for uncontracted integrals, the error arising from its use with contracted basis functions is expected to be larger than with basis sets containing fewer contracted functions.  The data in \figref{errors} show this, but they also show that the increase in error from the use of the \sqvl estimator is relatively small:  in the Def2-SVP/Def2-SVP/JK case with linear alkanes, the \sqvl estimator causes roughly a factor of two increase in the CADF-LinK error, and for water clusters the increase is even less.

\subsection{Speedups}

\begin{figure*}
  \centering
  \includegraphics[width=1.0\textwidth]{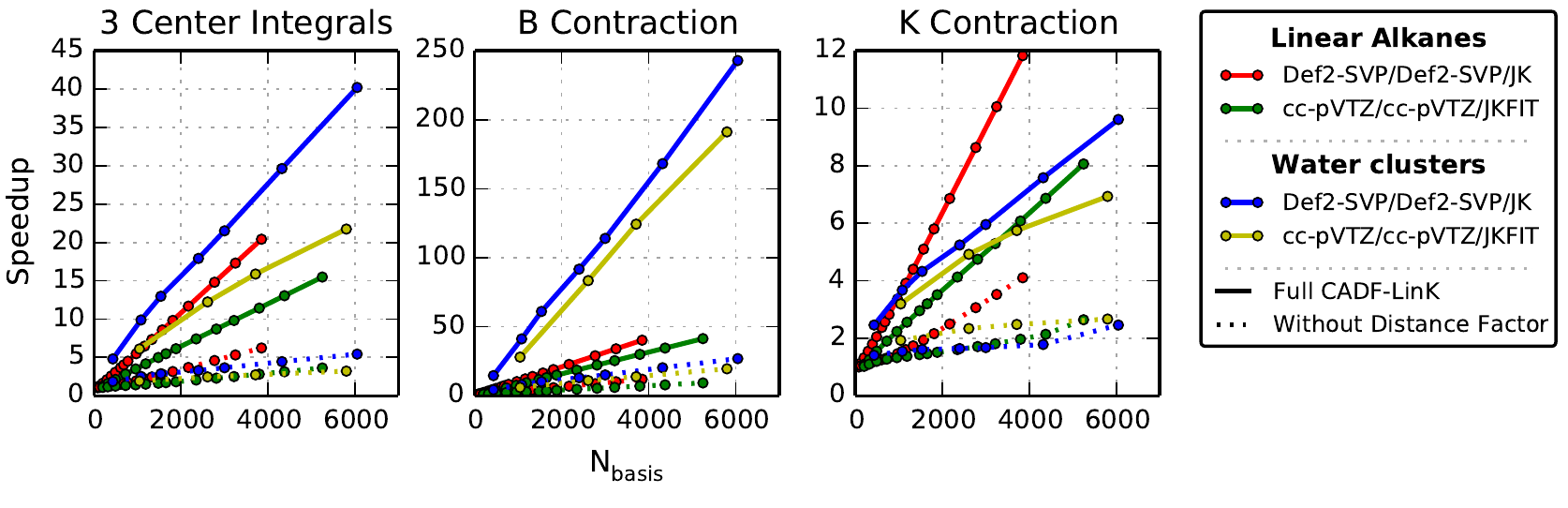}
  \caption{CADF-LinK speedups for the three cost metrics examined here relative to a CADF-based exchange build with Schwarz screening only.  The dotted lines show speedups for CADF-LinK without the distance-including \sqvl estimator. }
  \label{fig:speedup}
\end{figure*}

\figref{speedup} shows the speedups for the three cost metrics we examined relative to a CADF-based exchange build with Schwarz screening only---that is, the ratios of each cost metric for CADF exchange with Schwarz screening only to the cost metric for CADF-LinK for a given system size.  The most substantial speedups---as much as 250 times---are seen in the \exB contraction, which is expected given that both the inner and outer loops are restricted by CADF-LinK (by the \Lst B{\mu\la} and \Lst 3{\mu\la} lists, lines~\ref{al:blr} and \ref{al:3lr} respectively in \Algref{kb}), whereas the 3 center integrals and the \exK contraction are only restricted by one list each.  Indeed, the only significant additional restriction of the contraction in line~\ref{al:kcon2} relative to Schwarz screening comes from the outermost loop over $(\mu_a X_c)$ pairs (since the stronger concentricity restriction of the loop in line~\ref{al:k2out} is imposed even without the formation of \Lst B{\mu X}).  Consequently, the least substantial speedups are seen in the \exK contraction, though again this section is the least computationally intense of the three even for the largest computations we performed.  These data clearly show that the small increase in error (see \secref{err}) is, in most contexts, more than compensated for by a massive decrease in computational effort.

Another noteworthy trend in the data from \figref{speedup} is that for the first two cost metrics (3 center integrals and the \exB contraction), larger speedups were seen for three dimensional systems, while for the \exK contraction, larger speedups were seen for linear alkanes.  This is attributed to the more prominent role of the distance factor in the \Lst3{\mu X} definition compared to the \Lst B{\mu X} definition.  For \Lst3{\mu X}, the distance factor $\distf$ contributes directly to the thresholding for the inclusion of $\la$, whereas for \Lst B{\mu X}, the distance factors for all $\la$ connected by a P-junction to a given $\si$ contribute to the thresholding for the inclusion of $\si$.  Heuristically, the 3 center integrals cost metric is most influenced by the efficiency of \Lst3{\mu X}, and the \exK contribution is most influenced by the efficiency of \Lst B{\mu X} (the \exB cost metric is influenced by both).  Since we have already noted in \secref{err} that the three-dimensional structures are more susceptible to the effects of the distance-including estimator, it makes sense that more substantial gains would be seen with water clusters for the 3 center integrals and the \exB contraction and with linear alkanes for the \exK contraction.

The more interesting trend in \figref{speedup} is the contribution to the speedup from the distance-dependent integral screening.  For large water clusters with the Def2-SVP/Def2-SVP/JK basis pair, nearly tenfold speedups are observed compared to CADF-LinK without the \sqvl estimator.  This is remarkable, given that in the context of four-center ERI-based LinK the use of distance-dependent screening
results in speedups of about 2.0.\cite{Maurer:2012p144107}  This further demonstrates that there is more to be gained from screening three-center ERIs than from screening their four-center counterparts.

%%%%%%%%%%%%%%%%%%%%%%%%%%%%%%%%%%%%%%%%%%%%%%%%%%%%%%%%%%%%%%%%%%%%%%%%%%%%%%%%
%%%%%%%%%%%%%%%%%%%%%%%%%%%%%%%%%%%%%%%%%%%%%%%%%%%%%%%%%%%%%%%%%%%%%%%%%%%%%%%%
%%%%%%%%%%%%%%%%%%%%%%%%%%%%%%%%%%%%%%%%%%%%%%%%%%%%%%%%%%%%%%%%%%%%%%%%%%%%%%%%
%%%%%%%%%%%%%%%%%%%%%%%%%%%%%%%%%%%%%%%%%%%%%%%%%%%%%%%%%%%%%%%%%%%%%%%%%%%%%%%%
%%%%%%%%%%%%%%%%%%%%%%%%%%%%%%%%%%%%%%%%%%%%%%%%%%%%%%%%%%%%%%%%%%%%%%%%%%%%%%%%
%%%%%%%%%%%%%%%%%%%%%%%%%%%%%%%%%%%%%%%%%%%%%%%%%%%%%%%%%%%%%%%%%%%%%%%%%%%%%%%%
%%%%%%%%%%%%%%%%%%%%%%%%%%%%%%%%%%%%%%%%%%%%%%%%%%%%%%%%%%%%%%%%%%%%%%%%%%%%%%%%

\section{Conclusions}
\label{sec:conc}

We have presented a linear scaling algorithm for computing the exchange matrix using the concentric atomic density fitting approximation. Our algorithm has been shown to perform well for basis sets of all sizes, and we have carried out some of the largest triple- and quadruple-zeta basis computations ever to demonstrate this point.  Errors in absolute energies from the CADF-LinK approximation are substantially smaller than other sources of error and have been shown to grow linearly with basis set size.  Even for large basis sets, our method shows near linear scaling for systems of less than 1000 atoms, and for smaller basis sets the onset of linear scaling is even more rapid.  Not only does our algorithm serve as a highly efficient way to compute the exchange matrix in the context of Hartree-Fock and DFT methods, but also it offers a blueprint for the use of the concentric atomic density fitting in many-body methods.

%%%%%%%%%%%%%%%%%%%%%%%%%%%%%%%%%%%%%%%%%%%%%%%%%%%%%%%%%%%%%%%%%%%%%%%%%%%%%%%%
%%%%%%%%%%%%%%%%%%%%%%%%%%%%%%%%%%%%%%%%%%%%%%%%%%%%%%%%%%%%%%%%%%%%%%%%%%%%%%%%
%%%%%%%%%%%%%%%%%%%%%%%%%%%%%%%%%%%%%%%%%%%%%%%%%%%%%%%%%%%%%%%%%%%%%%%%%%%%%%%%
%%%%%%%%%%%%%%%%%%%%%%%%%%%%%%%%%%%%%%%%%%%%%%%%%%%%%%%%%%%%%%%%%%%%%%%%%%%%%%%%
%%%%%%%%%%%%%%%%%%%%%%%%%%%%%%%%%%%%%%%%%%%%%%%%%%%%%%%%%%%%%%%%%%%%%%%%%%%%%%%%
%%%%%%%%%%%%%%%%%%%%%%%%%%%%%%%%%%%%%%%%%%%%%%%%%%%%%%%%%%%%%%%%%%%%%%%%%%%%%%%%
%%%%%%%%%%%%%%%%%%%%%%%%%%%%%%%%%%%%%%%%%%%%%%%%%%%%%%%%%%%%%%%%%%%%%%%%%%%%%%%%

\section{Acknowledgements}

The research by DSH and EFV was supported by NSF grants CHE-0847295, CHE-1362655, and ACI-1047696, and a Camille Dreyfus Teacher-Scholar Award.  The research by DSH and HFS was supported by NSF grant CHE-1361178.
This work used resources of the National Energy Research Scientific Computing Center, which is supported by the Office of Science of the U. S. Department of Energy under Contract No. DE-AC02-05CH11231.

\end{document}